# $\Theta^+$ pentaquark In Multiquark Theories

## A. R. Haghpayma[†]

*Department of Physics, Ferdowsi University of Mashhad*

*Mashhad, Iran*

## Abstract

Now it is very difficult to calculate the whole hadron spectrum from first principles in QCD, under such a circumstance, various models which are QCD - based or incorporate some important properties of QCD were proposed to explain the hadron spectrum and other low- energy properties. In this paper we will explain and discus some features of them

QCD is believed to be the underlying theory of the strong interaction which has three fundamental properties: asymptotic freedom, color confinement, approximate chiral symmetry and its spontanous breaking; in high energy level QCD has been tested up to 0.01 level.

The behavior of QCD in the low energy is nonperturbative and the $SU_c(3)$ color group structure is non-abelian.

However, besides conventional mesons and baryons, QCD itself does not exclude the existance of the nonconventional states such as glueballs ( gg, ggg, ..... ) hybrid mesons ( $qq\bar{q}$g ), and other multi - quark states ( $qq\bar{q}q$, $qqqq\bar{q}$ ).

It is very difficult to calculate the whole hadron spectrum from first principles in QCD, under such a circumstance, various models which are QCD - based or incorporate some important properties of QCD were proposed to explain the hadron spectrum and other low- energy properties.

We will explain and discus some features of them:

1- Lattice Calculations

With the rapied development of new ideas and computing power, lattice gauge theory may provide the final solution to the spectrum problem in future.

Lattice QCD simulation may play a very important role eventually but right now, lattice simulation of pentaquarks by several groups has not converged yet, for example, one lattice calculation favors positive parity for pentaquarks[1], while two previous lattice simulations favor negative parity[2], some lattice simulations did not observe any bound pentaquark state in either I = 0 $J^P = 1/2^\pm$ or I = 1 $J^P = 1/2^\pm$ [3].

Soon after the first experimental reports of $\Theta^+$ the first lattice studies of it appeared[2].

There was confusion a both parity but both groups have found evidences for the KN threshold and for a state in the $uudd\bar{s}$ $J^P = 1/2^-$ channel. they also reported evidence for a state in the $1/2^+$ channel. but at a higher mass.

Although kentucky group[3] does not find a state in either parity channel.

For negative parity we have only one local source $qqqq\bar{q}$. but for positive parity we have eight local sources and the full correlation matrix with the eight local sources must be done to show the optimize overlape with possible positive parity state.

We may favour positive parity to negative one to resolve $\Theta^+$ width puzzle, for this reason we can use diquark ideas and and operatores[4] and do our calculations in the chiral limit ( hyperfine interactions between quarks ).

The goal of such ideas is diquark correlations[5,6] ( scaler diquarks - vector diquarks - .... )

Because of we are dealing with nearly massless quarks we are in the relativistic limit, but in lattice calculations the sources are classified according to their properties in the , non - relativistic-limit and by replacing Dirac quark fields by pauli fields.

The negative parity state observed in Ref [2,7] will turn out to have an enormous width and the state has exactly the same spin, color, and flavour wavefunction as K N in the S - wave and therefore should be very broad and there is a single $\overline{10}$ with either $J^P = 1/2^-$ or $J^P = 3/2^-$ and uncorrelated quarks in comparison to four octet with $J^P = 3/2^+$.

Thus the lattice calculations are at the beginning of their way.

2- Sum Rules.

The basic object in Sum Rule analysis is correlation functions for example:
$$\Pi(p) = i \int d^4x\, e^{ipx} \langle 0|T\{\eta(x)\bar{\eta}(0)\}|0\rangle, \quad (1)$$

Where $\eta(x)$ represents the interpolating field of the pentaquark under investigation.

Several interpolating fields has intended for a lattice search of $\Theta^+$ which are suitable for QSR analysis due to the QCD - based quark picture of pentaquark[8].

In ( j'w diquark model ) we have for $\eta(x)$:
$$\eta(x) = \left(\epsilon^{abd}\delta^{ce} - \epsilon^{abc}\delta^{de}\right)[Q_{ab}(D^\mu Q_{cd}) - (D^\mu Q_{ab})Q_{cd}]\gamma_5\gamma_\mu C\bar{s}_e^T, \quad (2)$$
in which $D^\mu = \partial^\mu - ig\lambda_l^i A^\mu l$ ( gluon exchange between diquarks and
$$Q^c(x) = \epsilon^{abc} Q_{ab}(x) = \epsilon^{abc}[u_a^T C\gamma_5 d_b](x). \quad (3)$$

C is charge conjugation matrix and a, b, c, are color index

In this model we have two scalar diquarks and there is electromagnetic interaction between $\bar{s}$ and diquarkes[9].

---

[†] e-mail: haghpeima@wali.um.ac.ir

The diquarks have a particularly strong attraction in the flavour antisymmetric $J^P=0^+$ channel, this attraction comes from hyperfine interactions between quarks, in a scalar diquark these interactions are flavour - spin and color - spin interactions thus there is exchanged gluons and mesons between two quarks in a diquark, the distance between two quarks is about 0.5 f and this calculations are chiral and relativistic. the two diquarks must be in a p- wave to satisfy Bose statistics, therefor the current contains a derivative to generate one unite angular momentum. the diquarks couple to a $3_c$ in colour to form the current $\eta(x)$, and the parity is positive.

In fact QSR calculations not pridict $\Theta^+$ pentaquark, but if it exist accomodate its mass.

It would be interesting to see if lattice calculations could confirm these findings. first lattice calculations exist[9] which, however, are based on different interpolating currents and whose results are not yet conclusive.

3- Larg $N_c$ QCD ($\chi$SM – SKM – non- Correlated P$\chi$QM)

The prediction of the mass, width and reaction channel of $\Theta^+$ from the chiral soliton model was the first prediction about it[41].

In this model there is a resonance S=1 J=1/2$^+$ at 1530Mev with a width less than 15 Mev and $\Theta^+$ is the lightest member of the anti- decuplet in the third rotational state of the chiral soliton model.

The mass of $\Theta^+$ has predicted by assuming that the N(1710) is a member of the anti- decuplet and by symmetry considerations of the model.

This will lead to a $\Xi^{--}$ pentaquark mass 210 Mev higher than that observed by NA49 collaboration[10].

Another reanalysis explain a fairly good description of both $\Theta^+$ and $\Xi^{--}$ masses in this model.

In $\chi$SM the narrow width of pentaquark coms from the cancellation between the coupling constan in the leading order, next - leading order and next - next - lading order large $N_C$ expansion[11].

The foundation of treatment of the pentaquarks is challenged by the $N_C$ formalism in Ref [12].

That prediction for a light collective $\Theta^+$ baryon state S=1 based on the collective quantization of chiral soliton models are shown to be inconsistent with large $N_C$ QCD.

Since collective quantization is legitimate only for excitations which vanish as $N \rightarrow \infty$.

In the large $N_C$ limit, the rotational degree of freedom decouples from the vibration mode, for example the nucleon octet and the $\Delta$ decuplet have a mass spliting of order O($1/N_C$) while the excitation energy of the vibration mode of the hedgehag is O(1).

In chiral soliton model the predictions of octet and decuplet multiplets are more reliable and in contrast the mass splitting between the anti- decuplet and octet is also O(1). this means that the rotation and vibration motions are not orthogonal.

They will mix each other, which invalidate the collective quantization of the rotational degree of freedom and the prediction of $\Theta^+$ properties may be fortuitous.

If $\Theta^+$ exist as a member of an antidecuplet large $N_C$ technique may be used to predict the existence of the other members of the same multiple thus this technique dosent predict the existence of $\Theta^+$ pentaquark, but accomodate its properties.

The investigation of the relationship between the bound states and SU(3) rigid rotator approaches to strangeness in the skyrme model has found that the exotic state may be an artifact of the rigid rotator approach to skyrm model for large $N_c$ and small $m$.

In skyrm model we can introduce a free quark lagrangian $\mathcal{L}_0$ and a lagrangian $\mathcal{L}_1$ contains meson fields $\pi^A$ exchange between quarks to have a chiral lagrangian.

The quarks have no mass and chiral symmetry breaking can generate their dynamical mass in nucleons:

$$\mathcal{L}_0 = \bar{q}(i\not\partial - M)q \quad (4)$$

$$\mathcal{L}_1 = \bar{q}[i\not\partial - M\exp(i\gamma^5\pi^A\lambda^a)]q \quad (5)$$

All of the quarks in a baryon are in a $\pi^A$ mean-field, thus one can say that baryons are solitons, for example solitons of the self consistent electrostatic fiel $U_c(\boldsymbol{x})$[13].

Then by use of the concept of SU(3) rigid rotator we can transform $U_c(\boldsymbol{x})$ into $U(\boldsymbol{x},t)$:

$$U(\boldsymbol{x},t) = A(t)U_c(\boldsymbol{x})A^\dagger(t), \quad (6)$$

Thus we are in the rotational mode of the $\chi$SM and there is no vibrational mode.

By expanding the total lagrangian in terms of momentum P and $1/N_c$ and considering low powers of P and high powers of $N_c$, we would have an effective lagrangian density at large $N_c$QCD expansion which is chiral and there is a chiral soliton field with its rotational mode; collective quantization of such rotational modes and relating of rotational tensor to $N_c$ and SU(3) degrees of freedom which leads to SU(3) multiplets are the next steps of $\chi$SM in the way to the prediction of the properties of the $\Theta^+$ and other parteners in their multiplets.

In this model there are a $\Theta^+$ singlet with I=0 J=1/2 (baryon octet) and a triplet with I=1, J=3/2[10,14], and a new anti - decuplet, identifying $P_{11}(1710)$ as member of anti - decuplet. predictions of this model for $\Theta^+$ are as follow:

M = 1.53 Gev, $\Gamma < 15$ Mev, I = 0, S = 1, $J^P = 1/2^+$ with $\Theta^+ \rightarrow K_0 p$ or $K^+ n$.

For baryons with spin 1/2 and + parity there is no N around, and a weak evidence for $\Sigma(1770)$, so Diakonov and petrov suggested a missing N around 1650 - 1690 Mev.

Noting is Exotic in the chiral soliton picture, baryons are solitons in the chiral meson field no baryon is exotic except that it has different quantum numbers compared to other baryons.

Assuming that chiral forces are essential in binding of quarks one gets the lowest baryon multiplets:

$$(8, 1/2), (10, 3/2), (\overline{10}, 1/2).$$

whose properties are relating by symmetry.

Effective field theories are not just models, they represent very general principles such as analyticity, unitarity, cluster decomposition of quantum field theory and the symmetries of the systems[15].

The chiral perturbation theory ($\chi$PT), for example, represents the low- energy behavior of QCD (at least in the meson sector).

Although baryons in Large -N limit behave like solitons, it is not clear in what theory they appear. A natural condidate is the $\chi$PT it seems that if baryons may appear as solitons, they should appear in the $\chi$PT, with infinitely many operators, but we may systematically expand the results with respect to the typical momentum scale P and keep a few operators at low- energy.

Thus one can say a general skyrm- witten soliton theory may be a systematic expansion of the soliton sector of the $\chi$PT with respect to N and P[16].

With a conventional $\chi$PT if we consider group theoretical clustering between quarks[6] and hyperfine QCD interactions between them, we have an correlated $\chi$PT between quarks or correlated P$\chi$QM. in other case we are working with non- correlated theories[17].

In fact, because of unconstraint number of degrees of freedom the uncorrelated five - body approaches lead to a larger number of degres of possible configurations of constituents than correlated ones.

On the other hand uncorrelated models cover a wide spectrum of possibilities for the possible pentaquark structure of the $\Theta^+$ baryon.

Moreover, in this treatment the quark fermi statistics can be imposed strictly, while in the correlated approaches it is only exactly fulfilled when the diquark is really a pointlike particle.

In an conventional model the baryons are described by their valence quarks as relativistic fermions $\psi(x)$ moving in an external field ( static potential ) :

$$V_{eff}(r) = S(r) + \gamma^0 V(r) \quad (7)$$

for example a confinement potential [18] with :

$$r = |\vec{x}| \quad S(r) = c\,r \quad V(r) = constant = V_0$$

The valence quark core is supplemented in the flavor SU(3) version by a clud of goldstone bosons ($\pi, K, \eta$) $\Phi_i(x)$ according to the chiral symmetry requirement.

and in addition by quantum flactuations of gluon field $A^a_\mu(x)$[19].

Treating also goldstone field as small flactuations around the valence quark core, one can drive the linearized effective lagrangian [18]:.

$$\mathcal{L}_{eff}(x) = \bar{\psi}(x)\left[i\not{\partial} - V_{eff}(r)\right]\psi(x) + \frac{1}{2}\sum_{i=1}^{8}[\partial_\mu \Phi_i(x)]^2 - \frac{1}{4}F^a_{\mu\nu}F^{a\,\mu\nu}$$
$$- \bar{\psi}(x)\left\{S(r)i\gamma^5\frac{\Phi(x)}{F} + g_s\gamma^\mu A^a_\mu(x)\frac{\lambda^a}{2}\right\}\psi(x) + \mathcal{L}_{\chi SB}(x) \quad (8)$$

where F = 88 Mev is the pion decay constant in the chiral limit[20] $g_s$ is the quark gluon coupling constant, $A^a_\mu(x)$ is the quantum component of gluon field and $F^a_{\mu\nu}$ is its conventional field strengh tensor.

$\hat{\Phi} = \sum_{i=1}^{8}\Phi_i \lambda_i = \sum_{P}\Phi_P \lambda_P$ the octet matrix of pseudo scalar mesons with $P = \pi^\pm, \pi^0, K^\pm, K^0, \bar{K}^0, \eta$.

The term $\mathcal{L}_{\chi SB}(x)$ contains the mass contributions both for quarks and mesons, which explicitly break chiral symmetry.

$$\mathcal{L}_{\chi SB}(x) = -\bar{\psi}(x)\mathcal{M}\psi(x) - \frac{B}{2}Tr[\hat{\Phi}^2(x)\mathcal{M}]. \quad (9)$$

here, $\mathcal{M} = diag\{m_u, m_d, m_s\}$ is the mass matrix of current quarks, $B = -\langle 0|\bar{u}u|0\rangle/F^2$ is the quark condensate constant.

Perturbation theory is formulated by the expansion respect to $\hat{\Phi}(x)/F \sim 1/\sqrt{N_c}$[21] and all calculations are performed at one loop or at order of accuracy $o(1/F^2, \hat{m}, m_s)$.

The explicit form of the ground state quark wave function is set up as :

$$u_0(\vec{r}) = \begin{pmatrix} g(r) \\ -if(r)\vec{\sigma}\cdot\hat{r} \end{pmatrix} Y^0_0(\hat{r})\chi_s \chi_f \chi_c, \quad (10)$$

and for an antiquark we have:

$$v_0(\vec{r}) = \begin{pmatrix} -l(r)\vec{\sigma}\cdot\hat{r} \\ ik(r) \end{pmatrix} Y^0_0(\hat{r})\chi_s \chi_f \chi_c, \quad (11)$$

By using of the interaction lagrangian , i.e the 4-th term of the total lagrangian of Eq (8).

The interaction lagrangian includes effects of the meson cloud and gluon corrections to the baryon by applying Wicks theorm with appropriate propagators for quarks, mesons and gluons. in the following equation we can find the energy shift of pentaquark valence particles interacting with pseudoscalar mesons and quantum gluon fields.

$$\Delta m_B = {}^B\langle\phi_0|\sum_{n=1}^{2}\frac{i^n}{n!}\int i\delta(t_1)d^4x_1\ldots d^4x_n T[\mathcal{L}_I(x_1)\ldots \mathcal{L}_I(x_n)]|\phi_0\rangle^B_c \quad (12)$$

and c referes to connected graphs only.

According to this calculations the contribution of pion- exchange flavour - spin interaction ( FS ) between two quarks and between a quark and an antiquark to the pentaquark mass shift is proportional to:

$$\langle B|\sum_{i<j}^{4}\sum_{a=1}^{3}\lambda^{(a)}_i\lambda^{(a)}_j\,\vec{\sigma}_i\cdot\vec{\sigma}_j\,|B\rangle, \quad (13)$$

and

$$\langle B|\sum_{i=1}^{4}\sum_{a=1}^{3}\lambda^{(a)}_i\lambda^{(a)}_5\,\vec{\sigma}_i\cdot\vec{\sigma}_5\,|B\rangle \quad (14)$$

and the contribution of gluon exchange color - spin interaction ( CS ) between two quarks to the pentaquark mass shift is proportional to: $\langle B|\sum_{i<j}^{4}\vec{\lambda}^C_i\cdot\vec{\lambda}^C_j|B\rangle$

If we discus on an correlated P$\chi$QM with explicit symmetry on configuration , this hyperfin interactions ( FS , CS ) would be considered on the four - quark subsystems which is used by many clustered quark models[22], and a special form of confining potential. in such models the model parameters , i.e the confining potential and the effective quark - gluon coupling, are set up and constrained such as to give a reasonable fit to mass shifts in multiplet ( octet or decuplet ) sector of conventional baryons. In fact in such models ($\chi$PQM) we keep only the terms which are of leading order in $N_c$ in the power counting expansions and our calculation is a systematic expansion in powers of $\delta m$, and we are unable to complete the perturbative calculation , because our calculational methods does not allows us to include representations beyond first a few ones.

From the restricted calculations , we see that the mixings among representations are large. it implies that the inclusion of more representations is important, and possible breakdown of the perturbative treatment.

Width calculations in such models shows differences between first and second order results and maybe second order calculations account for fine structures.

Thus it seems that we should complete the perturbative calculations including the mixing with an arbitrary many representations or using a completely different method[23].

4- Instanton - Liquid Model ( ILM ).

QCD instantons are known to produce deeply bound diquarks and it may be used as building blocks in the formation of multiquark states, in particular in dibaryons and pentaquarks. in this manner a new symmetry exists between states with the same number of -bodies -but different number of quarks appear, in particular the 3 - body pentaquarks can be naturally related to some excited baryons which leads to light dibaryon H as a limitation of this model.

This approach is based on correlation functions of nonlocal operators with 2 or 4 or 6 or ..... light diquarks and the effective interaction between diquarks have a repulsive core, due to pauli principle.

In this technique we start with the pairing first and follow small $N_c$ ideology a gainst large $N_c$ ideology in which we works on pseudoscalar mesons as clusters and uses of HF interactions $\chi$PQM belowe of the confining scale $\approx$ 200 Mev and restricting of the interactions between quarks to simple one- gluon and scalar meson exchange .

This HF interactions are between two quarks, whatever other quarks do quite differently, an instanton can serve one quark ( per flavor ) at time only, due to pauli principle for t' Hooft zero modes.

Thus the instantons significantly contribute to clustering ( both qq and q$\bar{q}$ )at small quark densities, however are much less able to do so for high density environment, and this creates quite significant repulsive interactions between diquarks, a typical repulsive scale 300 Mev with an approximately instanton radius $\rho = 0.35$ fm[24], reminiscent of nuclear core, and rather heavy multiquark states.

This interaction bound diquarks such as ( scalar - vector - tensor ) ones and because of similar mass and quantum numbers such as color charge, the diquarks may be considered on equal footing with constituent quarks.

Certain approximate symmetries then appear[25] relating states with different number of quarks but the same number of bodies.

Pentaquarks and dibaryons are in this model treated as 3 - body objects with two correlated diquarks plus an antiquark, are thus related to decuplet baryons; $\Theta^+ (1540) = (ud)\{ud\}\bar{s} = sss$ is an analogue of anti - $\Omega$ and is thus the top of the antidecuplet ( the conjugate of the decuplet )

C$\chi$PQM needs clustering justification and instanton induced interactions in clusters can help to this problem.

The instantons, strong fluctuations of gluon fields in the vacuum, play a crucial role in the realization of spontaneous chiral symmetry breaking in QCD.

The instantons induce the t' Hooft interaction between the quarks which has strong flavor and spin dependence, a behavor which explains many features observed in the hadron spectrum and in hadronic reactions[26].

For example this interaction prefer scalar diquarks to other ones with an interaction strength comparable to the pion channel pauli - Gursey symmetry and only one - half weaker in the realistic $N_c = 3$ case[25], thus favors correlated quark models by governing instanton induced interaction between quarks at intermediate distances, i. e. $r \approx \rho_c \approx 0.3\,f\rho_c$ is the average instanton size in the QCD vacuum[27] and is much smaller than the large confinement region R=1 fm.

The most important non- perturbative instanton induced interaction among several approaches[28] is the multiquark t' Hooft interaction which arises from the quark zero modes in the instanton field[29].

In the limit of small instanton size the effective two and three - body point - like interactions are[28] :

$$\mathcal{H}_{eff}^{(2)}(r) = -V_2 \sum_{i \neq j} \frac{1}{m_i m_j} \bar{q}_{iR}(r) q_{iL}(r) \bar{q}_{jR}(r) q_{jL}(r) \left[ 1 + \frac{1}{32}(\lambda_u \lambda_d + \text{perm.}) + \frac{9}{32}(\vec{\sigma_u} \cdot \vec{\sigma_d} \lambda_u^a \lambda_d^a + \text{perm.}) \right] + (R \longleftrightarrow L), \quad (15)$$

$$\mathcal{H}_{eff}^{(3)}(r) = -V_3 \prod_{i=u,d,s} \bar{q}_{iR}(r) q_{iL}(r) \left[ 1 + \frac{3}{32}(\lambda_u^a \lambda_d^a + \text{perm.}) + \frac{9}{32}(\vec{\sigma_u} \cdot \vec{\sigma_d} \lambda_u^a \lambda_d^a + \text{perm.}) - \frac{9}{320} d^{abc} \lambda^a \lambda^b \lambda^c (1 - 3(\vec{\sigma_u} \cdot \vec{\sigma_d} + \text{perm.})) - \frac{9 f^{abc}}{64} \lambda^a \lambda^b \lambda^c (\vec{\sigma_u} \times \vec{\sigma_d}) \cdot \vec{\sigma_s} \right] + (R \longleftrightarrow L)$$

where m is the effective quark mass. $q_{R,L} = (1 \pm \gamma_5) q(x)/2$ (16)

$$\mathcal{H}_{eff}^{(2) q \bar{q}} = -\sum_{i \neq j} \frac{a}{m_i m_j} \left[ 1 - \frac{3}{32}(\lambda_u^a \lambda_s^a + \text{perm.}) + \frac{9}{32}(\vec{\sigma_u} \cdot \vec{\sigma_{\bar{s}}} \lambda_u^a \lambda_s^a + \text{perm.}) \right], \quad (17)$$
and $\lambda_{\bar{q}} = -\lambda^*, \sigma_{\bar{q}} = -\sigma^*$

The perturbative OGE hyperfine interactions (HFI) between quark and confinement in this clusters (CS, FS) are the residual interactions and one can include them to this (ILI) interactions.

Once calculating the mass shift of hadron states by means of these interactions one can use an conventional quark model mass formula for obtaining the masses of hadron multiplet [30].

In order to capture correctly the physics of QCD between the confinement scale and the chiral symmetry breaking scale one can use of the diquark chiral effective theory[31].

The relevant degrees of freedom of diquark chiral effective theory are constituent quarks, diquarks, gluons, and pions.

When the diquarks are absent or infinitly heavy, the effective lagrangian should reduce to the chiral quark effective theory of Georgi and Manuher.

In such models we can calculate QCD-based quark - diquark bound state energy insted of consideration $l=1$ angular momentum between them in a scattering potential.

Since the diquark masses (e.x.scalar or tensor, ..... ) are smaller than the constituents, they are stable against decay near mass shell, in such a configuration, the diquarks are nearby and tunneling of one of the quarks between the two diquarks may take place.

4-1-Decay widths.

In $\Theta^+ \to K^+ N$ decay a d quark tunnels from a diquark ud to the other diquark to form a nucleon udd and an off- shell u quark, which is annihilated by the anti-strange quark . (if u were to tunnel, the decay is to $K^0 P$ with a comparable decay width.).

The decay width is therefor given as:
$$\Gamma = \lim_{v \to 0} \sigma(\bar{s} + \phi_{ud} + \phi_{ud} \to K^+ + n) v \, |\psi(0)|^2, \quad (18)$$

where $v$ is the velocity of $\bar{s}$ in the rest frame of the target diquark and $\psi$ is the 1S wave function of the quark - diquark inside the pentaquark.

The differential cross section for the annihilation process is then:
$$d\sigma = \frac{(2\pi)^4 |\mathcal{M}|^2}{4 \sqrt{(p_1 \cdot p_2)^2 - m_s^2 M_{ud}^2}} \, 4 e^{-2S_0} \, d\Phi(p_1 + p_2; k_1, k_2), \quad (19)$$

in which $e^{-2S_0}$ is tunneling probability.

If we insert annihilation amplitude $|\mathcal{M}|$ and phase space $d\Phi$ and integrate over the $d\Phi$ and taking $v \to 0$ we find:
$$\Gamma_{\Theta^+} \simeq 5.0 \, e^{-2S_0} \frac{g^2 g_A^2}{8 \pi f_K^2} |\psi(0)|^2. \quad (20)$$

by using the WKB approximation for $e^{-2S_0}$:
$$e^{-S_0} = \langle n | T e^{i \int d^4 x \mathcal{L}_{int}} | d, \varphi_{ud} \rangle \approx e^{-\Delta E r_0}, \quad (21)$$
where $\Delta E = (m_u + m_d) - M_{ud}$

and for $r_0$ we have:
$$M_{\Theta^+} = 2 M_{ud} + m_{\bar{s}} + \frac{2}{M_{ud} r_0^2}, \quad (22)$$

where the third contribution is the rotational energy of diquarks in a p - wave.

The 1 S wave function of the quark - diquark at the origin can be written as:
$$\psi(0) = \frac{2}{a_0^{3/2}} \frac{1}{\sqrt{4\pi}}, \quad (23)$$

where $a_0$ is the Bohr radius of the quark - diquark bound state.

Assuming they are non- relativistic we get by the dimentional analysis $a_0 \simeq (2 \bar{m} B)^{-1/2}$.

where $\bar{m} = 250$ Mev is the reduced mass and B is binding energy of quark - diquark bound state.

Taking $B = 100 \sim 200$ Mev , comparable to the pentaquark binding energy, $g^2 = 3.03$ and $g_A = 0.75$ from the quark model , one can find the $\Theta^+$ width .

5- Susy Model.

According to QCD, an approximate dynamical supersymmetry exist between an antiquark and a diquark, the first person to point out a supersymmetry between antiquarks and diquark was Miyazawa[32].

The point is that both a diquark and an antiquark belong to an antidecuplet of $SU(3)$, and to the first approximation the interaction of QCD depends only on color.

Thus using the framework of the constituent quark model and according to QCD this supersymmetry ( diquark - antiquark ) exist which pointed out by Catto and Gürsey[33] they mensioned that mesons and baryons have Regge trajectories with approximately the same slop and it is because of existing this kind of supersymmetry. but this supersymmetry is brocken because of different mass, spin, and size of a diquark and its partener antiquark, the mass of an antiquark is smaller than diquark .

Although the mass of a diquark is smaller than the mass of its constituent quarks because of the interactions between the quarks which are QCD based.

No one knows how to reduce the mass of a diquark and reach the mass of an antiquark (unknown mecanism).

Because of different spins of a diquark and an antiquark they would have different QCD-based interactions with other particles,

A diquark may be almost the size of a hadron, a current quark is assumed to be pointlike, however a constituent quark is certainly not pointlike and there are cloud of gluons and quark - antiquark pairs around it.

Now one can consider a conventional anti-baryon $\bar{\Lambda} = \bar{u} \bar{d} \bar{s}$ and replace $\bar{u}$ and $\bar{d}$ antiquarks by $[ds]$ and $[su]$ diquarks and reach to pentaquark $dssu\bar{s}$ ).

By assuming that diquarks are in $\bar{3}_c$ in SU(3) - color, we have for example: ( $J^{'}W$ model ).

In the J W diquark model diquarks are in $\bar{3}_f$ in SU(3) - flavour and this replacing are in both color - flavour space.

(F S) interactions lower the mass of diquarks and in this model the difference between the mass of a diquark and an antiquark is smaller than the configuration in which diquarks are not in $\bar{3}_f$ configuration in SU(3)- flavor space.

If one consider $m_u \simeq m_d \simeq m_s = 1/3 \, m_N$, and consider pentaquark ( $dssu\bar{s}$ ) as a $\Theta^+$ ( $uudd\bar{s}$ ), its mass would be greater than experimental limits [34].

In the model there is noting to prevent of a very broad width for $\Theta^+$ and there is another unknown mechanism to be discovered to explain the experimental results for the width of $\Theta^+$.

6. constituent quark model.

Pentaquark baryons may be pure exotic or Crypto- exotic the pure exotic states can easily be identified by their unique quantum numbers, but the crypto - exotic states are hard to be identified as their quantum numbers can also be generated by three - quark states. therefore, it is crucial to have careful analyses for their decay channels.

Quark models have provided a cornerstone for hadron physics and one can studying the structure of pentaquark baryons in a naive constituent quark model[35].

In Ref [22] karliner and lipkin suggested a triquark - diquark model where , for example, $\Theta^+$ is a system of $(ud) - (ud\bar{s})$.

In Ref [6] jaff and wilczek presented a diquark - diquark- antiquark model so $\Theta^+$ is $(ud) - (ud) - \bar{s}$, in this model they also considered the mixing of the pentaquark antidecuplet with the pentaquark octet, which makes it different from the SU(3) soliton models where the octet describes the normal three- quark baryon octet.

More predictions based on quark models can be found in Ref [35 , 36].

References

[1] T . - W. Chiu, T.-H.Hsieh, hep - ph/0403020

[2] F. Csikor et al.,hep - lat/0309090;S. Sasaki, hep-lat/0310014.

[3] N.Mathur et al., hep-ph/0406196.

[4] R.Jaffe and F.Wilczek,arXiv:hep-ph/0401034.

[5] S. Nussinov, hep-ph/0307357;R. A. Arndt,I.I.Strakovsky, R. L. Workman, Phys. Rev. C 68, 042201 (R) (2003); J.Haidenbauer and G. Krein, hep-ph/0309243;R. N. Cahnand G. H. Trilling, hep-h/0311245.

[6] R.Jaffe and F.Wilczek,Phys.Rev.Lett.91, 232003 (2003).

[7] N.Mathur et al, arXiv:hep-ph/0406196.
[8] L. Ya. Glozman, ariv:hep-ph/0308232.
[9] Markus Eidemuller arXiv:hep-ph/0404126.
[10] D. Diakonov et al., Z. Phys. A 359, 305 (1997).
[11] M.Praszałowicz,Phys.Lett.B 583 , 96 (2004) [arXiv:hep-ph/0311230].
[12] R.L.Jaffe,Eur.Phys.J.C 35 , 221 (2004) [arXiv:hep-ph/0401187].
[13] E.Witten,Nucl.Phys.B 223 , 433 (1983).
  Wess and B. Zumino, Phys. Lett. B37, 95 (1971).
  E. Witten, Nucl. Phys. B 223 , 422 (1983).
[14] Borisyuk , arXiv:hep-ph/0404126.
[15] S.Weinberg,PhysicaA 96 , 327 (1979).
[16] Koji Harada . et al, arXiv:hep-ph/0410145.
[17] T. Inoue et al , arXiv:hep-ph/0407305.
[18] T.Inoue,V.E.Lyubovitskij,T.GutscheandA.Faessler,
  [arXiv:hep-ph/0404051].
[19] H.Leutwyler,Nucl.Phys.B 179, 129 (1981); D. Diakonov and V. Y. Petrov,
  Nucl. Phys. B 245 , 259 (1984); G. V. Efimov and M. A. Ivanov,
  *The Quark Confinement Model of Hadrons* (IOP Publishing, Bristol & Philadelphia, 1993).
[20] J.Gasser,M.E.SainioandA.Svarc,Nucl.Phys.B 307, 779 (1988).
[21] V. E. Lyubovitskij, T. GutscheandA. Faesslerw, Phys. Rev. C 64 , 065203,
  (2001)[arXiv:hep-ph/0105043].
[22] M.KarlinerandH.J.Lipkin,arXiv:hep-ph/0307243.
[23] H.YabuandK.Ando,Nucl.Phys.B 301 , 601 (1988).
[24] Edward Shuryak , arXiv:hep-ph/0505011.
[25] E.ShuryakandI.Zahed, hep-ph/0310270
[26] T.Sch¨aferandE.V.Shuryak,Rev.Mod.Phys. **70** (1998) 1323.
  D. Diakonov, Prog. Par. Nucl. Phys. **51** (2003) 173.
  A.E.Dorokhov, N.I.Kochelev and Yu.A. Zubov, Int. J. Mod. Phys. **8A** (1993) 603.
[27] E.V.Shuryak,Nucl.Phys. **B203**, 93; 116; 140 (1982).
[28] A.E.DorokhovandN.I.Kochelev,PreprintJINR-E2-86-847 (1986), available from
  KEK library.
  A.E. Dorokhov, N.I.Kochelev, Yu.A. Zubov, Yad. Fiz. **50**(1989) 1717 (Sov. J. Nucl.
  Phys. **50** (1989) 1065), Z. Phys. **C65** (1995) 667; N.I. Kochelev, JETP Lett.**70**
  (1999) 491.
  S. Takeuchi and M. Oka, Phys. Rev. Lett. **66**(1991) 1271.
[29] M.A.Shifman,A.I.Vainshtein,A.I.Zakharov,Nucl.Phys. **B163**, 43 (1980).
[30] E.V.ShuryakandJ.L.Rosner,Phys.Lett. **B218** (1989) 72.
[31] Deog Ki Hong et al, arXiv:hep-ph/0403205.
[32] H.Miyazawa,Prog.Theor.Phys.**36**, 1266 (1966).
[33] S.CattoandF.G¨ursey,NuovoCimento**86**, 201 (1985).
[34] D.B. Lichtenberg, arXiv:hep-ph/0406198.
[35] R.Bijker,M.M.Giannini,andE.Santopinto,hep-ph/03 10281.
  Fl. Stancu and D. O. Riska, Phys. Lett. B 575 , 242 (2003).
  C. E. Carlson, C. D. Carone, H. J. Kwee, and V. Nazaryan, Phys. Lett. B 573,
  101 (2003).
  C. E. Carlson, C. D. Carone, H. J. Kwee, and V. Nazaryan, Phys. Lett. B 579,
  52 (2004).
  S. M. Gerasyuta and V. I. Kochkin, hep-ph/0310227.
  N. I. Kochelev, H. J. Lee, and V. Vento, hep-ph/0404065.
[36] A. R. Haghpayma, hep-ph/0606162.